\documentclass[a4paper,fleqn,preprint,numbers]{elsarticle}

\biboptions{sort&compress}

\usepackage[utf8]{inputenc}
\usepackage[T1]{fontenc}

\usepackage{graphicx}
\usepackage{amsmath}
\usepackage{amssymb}
\usepackage{lineno}
\usepackage{subfig}
\usepackage{mathtools}
\usepackage{bbm}
\usepackage{tikz}
\usepackage{booktabs}
\usepackage{float}
\usepackage{stfloats}
\usepackage{placeins}
\usepackage[hidelinks]{hyperref}
\usepackage{glossaries}

\newacronym{fos}{FOS}{fractional-order system}
\newacronym{elbo}{ELBO}{evidence lower bound}
\newacronym{vb}{VB}{variational Bayes}
\newacronym{eis}{EIS}{electrochemical impedance spectroscopy}
\newacronym{drt}{DRT}{distribution of relaxation times}
\newacronym{ecm}{ECM}{equivalent circuit model}
\newacronym{sofc}{SOFC}{solid-oxide fuel cell}
\newacronym{soec}{SOEC}{solid-oxide electrolyser cell}
\newacronym{uot}{UOT}{unbalanced optimal transport}
\newacronym{ot}{OT}{optimal transport}
\newacronym{cdf}{CDF}{cumulative distribution function}
\newacronym{pdf}{PDF}{probability density function}
\newacronym{cdrt}{CDRT}{cumulative distribution of relaxation times}
\newacronym{wd}{WD}{Wasserstein distance}
\newacronym{nn}{NN}{neural network}
\newacronym{cnn}{CNN}{convolutional neural network}
\newacronym{fcnn}{FCNN}{fully connected neural network}
\newacronym{dpp}{DNN}{deep neural network}
\newacronym{rf}{RF}{receptive field}
\newacronym{cae}{CAE}{convolutional autoencoder}
\newacronym{vae}{VAE}{variational autoencoder}
\newacronym{bigru}{BiGRU}{bidirectional gated recurrent unit}
\newacronym{soh}{SOH}{state of health}
\newacronym{icae}{ICAE}{improved convolution autoencoder}
\newacronym{tcn}{TCN}{temporal convolutional network}
\newacronym{kl}{KL}{Kullback--Leibler}
\newacronym{mae}{MAE}{mean absolute error}
\newacronym{rmse}{RMSE}{root mean squared error}
\newacronym{rbf}{RBF}{radial basis function}

\usepackage[nomargin,inline,marginclue,draft,multiuser]{fixme}
\fxsetup{theme=color,mode=multiuser}
\FXRegisterAuthor{pb}{apb}{\color{red}PB}
\FXRegisterAuthor{mb}{amb}{\color{blue}MB}
\FXRegisterAuthor{zg}{arf}{\color{brown}ZG}
\FXRegisterAuthor{zang}{zng}{\color{green}ZanG}
\FXRegisterAuthor{rs}{ars}{\color{teal}R2}

\usetikzlibrary{bayesnet}
\usetikzlibrary{positioning,arrows.meta,shapes.misc}
\usetikzlibrary{calc}

\definecolor{PastelPurple}{RGB}{200,180,240}
\definecolor{PastelGreen}{RGB}{180,225,180}
\definecolor{PastelYellow}{RGB}{255,245,180}
\definecolor{PastelPink}{RGB}{250,200,210}
\definecolor{PastelRed}{RGB}{240,170,170}
\definecolor{PastelBlue}{RGB}{190,215,255}
\definecolor{PastelOrange}{RGB}{255,220,180}

\journal{International Journal of Hydrogen Energy}

\begin{document}

\begin{frontmatter}

\title{Physics-informed distribution of relaxation times estimation and latent-space condition monitoring of solid oxide fuel and electrolysis cells from electrochemical impedance spectroscopy}

\author[1,2]{{\v Z}an Gorenc\corref{cor1}}
\ead{zan.gorenc@ijs.si}

\author[1,2]{{\v Z}iga Gradi{\v s}ar}
\ead{ziga.gradisar@ijs.si}

\author[3]{Felix Mütter}
\ead{felix.muetter@tugraz.at}

\author[3]{Vanja Suboti\'{c}}
\ead{vanja.subotic@tugraz.at}

\author[1,4]{Pavle Bo{\v s}koski}
\ead{pavle.boskoski@ijs.si}

\cortext[cor1]{Corresponding author}

\affiliation[1]{
    organization={Jo{\v z}ef Stefan Institute},
    addressline={Jamova cesta 39},
    city={Ljubljana},
    postcode={1000},
    country={Slovenia}
}

\affiliation[2]{
    organization={Jo{\v z}ef Stefan International Postgraduate School},
    addressline={Jamova cesta 39},
    city={Ljubljana},
    postcode={1000},
    country={Slovenia}
}

\affiliation[3]{
    organization={Institute of Thermal Engineering,
        Graz University of Technology},
    addressline={Inffeldgasse 25/B},
    city={Graz},
    postcode={8010},
    country={Austria}
}

\affiliation[4]{
    organization={Faculty of Information Studies in Novo mesto},
    addressline={Ljubljanska cesta 30},
    city={Novo mesto},
    postcode={8000},
    country={Slovenia}
}

\begin{abstract}
Estimating the \gls{drt} from \gls{eis} is an ill-posed inverse problem that is highly sensitive to regularisation choices. 
We propose a physics-informed convolutional autoencoder that estimates \glspl{drt} directly from \gls{eis} data without spectrum-specific tuning.
A discretised relation between impedance and the \gls{drt} is embedded in the training process, constraining the network to produce impedance-consistent distributions.
The model resolves overlapping relaxation processes in synthetic two-ZARC spectra and accurately reconstructs measurements from three independent solid oxide fuel and electrolysis cell datasets, with range-normalised errors below 1.1\%. 
Decoder-probe analysis shows that the learned latent representation is organised according to relaxation timescale. 
Distances in this latent space capture operating changes, hydrogen-shortage events, and long-term degradation. 
The same lightweight architecture is applied across all datasets without modification, providing consistent \gls{drt} estimation and an interpretable basis for condition monitoring.
\end{abstract}


\end{frontmatter}


\glsresetall

\section{Introduction}\label{sec:intro}


In electrochemistry, characterization of a device is often performed via \gls{eis}, which provides the impedance or frequency response of the system.
While \gls{eis} is a powerful diagnostic method, interpreting raw complex-valued impedance across many frequencies remains challenging.

To deconvolve the raw \gls{eis} spectrum, a widely known method is to fit an \gls{ecm} to the data.
\Glspl{ecm} require quite specific prior assumptions for the identification, which is a particularly acute problem for fractional-order systems where no systematic structural identification strategy exists.
Moreover, as discussed by \citet{ZNIDARIC2021117101}, impedance based analysis is subject to uncertainty arising from measurement noise, operating variability, and modelling assumptions, which can further obscure parameter identifiability and interpretation.

To overcome these limitations, \gls{drt} has emerged as a widely adopted method for analysing \gls{eis} data~\cite{barsukov2012electrochemical}.
\Gls{drt} decomposes the impedance spectrum into contributions associated with different relaxation time constants, offering a physically interpretable representation that replaces discrete circuit selection with a continuous distribution over an implicit Voigt structure~\cite{10.1021/ja01847a013}.
The relation between \gls{drt} and impedance \(Z(j\omega)\) is given by~\cite{10.1016/j.jpowsour.2023.233845}
\begin{equation}
    Z(j\omega) = \int_{-\infty}^{\infty} \frac{g(\log \tau)}{1 + j \omega \tau} \, \mathrm{d}\log \tau,
    \label{eq:drt_original}
\end{equation}
where \(Z(j\omega)\) denotes the impedance, \(\omega\) the angular frequency, \(\log \tau\) the logarithmic relaxation time, and \(g(\log \tau)\) the \gls{drt} function.

The motivation for using \gls{drt} is twofold.
First, the kernel \(1/(1 + j \omega \tau)\) represents the impedance of a simple (parallel) RC element, which relaxes exponentially with time constant \(\tau\) under an impulse perturbation.
Thus, \(g(\log \tau)\) can be interpreted as a superposition of many such elementary processes, revealing the dominant relaxation modes of the system~\cite{10.1021/ja01847a013}.
This is analogous to representing complex dielectric relaxation behaviour using a distribution of Debye relaxation times~\cite{10.1021/ja01847a013}.
Second, unlike complex impedance, \gls{drt} maps each frequency to a single scalar value, simplifying visualisation and interpretation.
Note that \eqref{eq:drt_original} omits explicit resistive and inductive terms to focus solely on the relaxation behaviour.

However, extracting the \gls{drt} from measured impedance data is an ill-posed inverse problem.
Small variations or noise in the impedance measurements can produce significantly different \glspl{drt} that nonetheless yield similar Nyquist curves.
To obtain physically meaningful results, regularisation is typically introduced~\cite{tikhonov}.
The choice of regularisation strongly influences the resulting \gls{drt}, making the analysis sensitive to the selected method.
Regularisation may be chosen based on self-consistency criteria~\cite{saccoccio2014optimal} or cross-validation approaches~\cite{maradesa2023selecting}.
To avoid hand-tuned regularisation, data-driven approaches offer an attractive alternative~\cite{liu2020gaussian,10.1149/1945-7111/ab631a}.

Physics-based neural networks provide a way of employing the powerful approximation mechanisms of neural networks while at the same time constraining the model output to satisfy known physical relationships~\cite{10.1016/j.jcp.2018.10.045,10.48550/arxiv.2201.05624}.
This concept is directly transferable to the estimation of the \gls{drt}.
In this setting, the unknown function $g(\log\tau)$ is represented by a neural network while being constrained such that its integral satisfies the impedance relation in \eqref{eq:drt_original}.

A notable example is the approach proposed by \citet{10.1149/1945-7111/ab631a}, which estimates \glspl{drt} directly from individual \gls{eis} spectra using a Deep Prior neural network.
The method solves the integral in \eqref{eq:drt_original} through a Fredholm approximation and proves to be effective even in the presence of noise and overlapping relaxation processes.
However, because a separate network is trained for each spectrum, the implicit regularisation, induced by the neural network, varies from measurement to measurement, making it perilous to attribute differences in estimated \glspl{drt} to genuine electrochemical changes rather than fitting artefacts.

Recent studies have demonstrated the effectiveness of learning shared latent representations from \gls{eis} data.
Autoencoder architectures have been successfully applied to battery diagnostics, where latent variables extracted from impedance spectra capture information related to degradation and improve \gls{soh} estimation~\cite{Obregon2023,Liu2024,Lou2025}.
Several works have also shown that transforming impedance spectra into \gls{drt} representations via \gls{rbf} deconvolution prior to learning improves interpretability and predictive performance by separating overlapping electrochemical processes into distinct relaxation features~\cite{Kim2025,Li2025}.
However, these approaches either treat the latent space as a black-box feature extractor or rely on a separate deconvolution step.
None embed the \gls{drt} integral relation directly into the learning objective.

We propose a physics-constrained \gls{cnn} framework that estimates \glspl{drt} from multiple \gls{eis} spectra simultaneously.
In contrast to approaches that fit a separate model to each spectrum, our method learns a shared latent representation across the entire dataset, enforcing consistent regularisation and enabling direct comparison of the resulting \glspl{drt}.
The learned latent space provides a compact, interpretable basis for analysing temporal evolution and condition-dependent behaviour, while decoder probe analysis reveals that distinct latent channels correspond to specific relaxation time regions.
We validate the framework on three independent solid oxide cell datasets, comprising steam electrolysis (603 spectra), co-electrolysis (150 spectra), and reversible operation (1248 spectra), demonstrating that predicted \glspl{drt} reproduce degradation trends established by classical analysis.

\section{Convolutional formulation of the inverse problem}
Solving for $g(\log \tau)$ can be recast as a convolution, given in general by
\begin{equation}
(h*f)(t)
=
\int_{-\infty}^{\infty}
h(\tau)\,f(t-\tau)\,\mathrm d\tau .
\end{equation}
The relation between \gls{drt} and impedance can be thus written as a convolution
\begin{align}
\hat Z(\log \hat\tau)
&=
\int_{-\infty}^{\infty}
\frac{
g(\log\tau)\,\mathrm d\log\tau
}{
1+j\exp\!\left[-(\log\hat\tau-\log\tau)\right]
}
\nonumber\\
&=(g*f)(\log\hat\tau).
\end{align}
where $\hat Z (\log \hat \tau) = Z(j \omega) $, $\hat \tau = 1/\omega$
and  $f(x)= 1/\left(1+j \exp\left(-x\right)\right)$.

Convolution theorem states that Fourier transform of a convolution of two functions is the product of their Fourier transform~\cite{oppenheim2013}, i.e. $\mathcal{F}\{\hat Z\} = \mathcal{F}\{g\} \mathcal{F}\{f\}$.
Rearranging the equations we get $\mathcal{F}\{g\}  = \mathcal{F}\{\hat Z\}  \frac{1}{\mathcal{F}\{f\} }$,
which we can again write out as a convolution
\begin{align}
	g (\log \tau) = \left(\hat Z * h\right)(\log \tau),
\end{align}
where $h = \mathcal{F}^{-1}\left\{ \frac{1}{\mathcal{F}\{f\} }\right\}$ can be seen as a filter that filters out \gls{drt} from impedance.

Since filters can be decomposed to what is know as cascade form~\cite[p.~61]{cascade}
\begin{align}
(\hat Z*h)(\log\tau)
&\simeq
\Bigl(
((\hat Z*h_1)*h_2)
*h_3
\cdots
\Bigr)
(\log\tau),
\label{eq:cascade}
\end{align}
where filters $h_1,h_2,h_3, \dots$ can perform several different tasks such as denoising~\cite{wiener}, compression-decompression~\cite{cascade} and in the end return a regularized solution~\cite[Chapter 6]{Hansen2006Deblurring}.

The cascade representation in~\eqref{eq:cascade} suggests that the inverse \gls{drt} problem can be approximated by a sequence of filtering operations.
Since convolutional layers perform precisely such operations, \glspl{cnn} are a natural choice for this task.
This motivates the encoder-decoder architecture described in next section, where convolutional layers learn the cascade of filters directly from data.

\section{Physics-informed convolutional neural network framework}
The proposed framework employs a \gls{cnn} to estimate \glspl{drt} from \gls{eis} data.
\Glspl{cnn} exploit local correlations through trainable kernels that slide along the frequency axis, reducing the number of parameters while capturing frequency-dependent features within a defined receptive field~\cite{oshea2015introductionconvolutionalneuralnetworks}.
Each kernel operates within a defined \gls{rf} and captures relevant spectral patterns while substantially reducing the number of trainable parameters compared to fully connected architectures.
The feature extraction process is controlled by the kernel size, stride, and padding, which determine the output dimensionality and receptive field~\cite{oshea2015introductionconvolutionalneuralnetworks}.

\begin{figure}[h]
    \centering
    \includegraphics{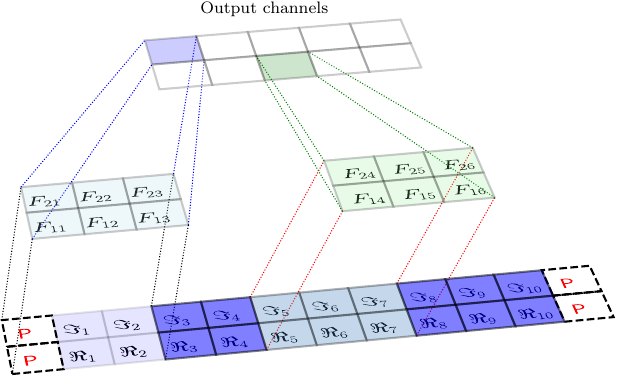}
    \caption{\Gls{cnn} first layer schematic representation for kernel size 2, padding 1 and stride 1.}\label{fig:first_layer}
\end{figure}

In this work, the real and imaginary components of the impedance are treated as two input channels (\figurename~\ref{fig:first_layer}).
Each kernel spans both channels and slides along the frequency axis, allowing the network to jointly learn local patterns across both components of the complex impedance spectrum.

Within the proposed framework, the \gls{cnn} does not predict impedance directly.
Instead, it estimates the \gls{drt}, \(g(\log\tau)\), which is subsequently inserted into a discretised form of~\eqref{eq:drt_original}.
To compute the corresponding impedance, the continuous integral is numerically approximated over a grid of logarithmic relaxation times using the trapezoidal rule,
\begin{equation}
    \int_a^b f(x)\,dx
    \approx
    \sum_{k=1}^{N}
    \frac{f(x_{k-1})+f(x_k)}{2}
    \Delta x_k,
    \label{eq:trapz}
\end{equation}
where \(f(x)\) corresponds to the integrand of~\eqref{eq:drt_original} and \(x_k\) denotes sampled values of \(\log\tau\).

\begin{figure}[h]
    \centering
    \resizebox{\linewidth}{!}{%
        \begin{tikzpicture}[
                node distance=1.4cm,
                every node/.style={font=\small},
                box/.style={
                        draw,
                        rounded corners,
                        align=center,
                        minimum width=3.0cm,
                        minimum height=1.1cm,
                        fill=#1!30
                    },
                smallbox/.style={
                        draw,
                        rounded corners,
                        align=center,
                        minimum width=2.7cm,
                        minimum height=0.9cm,
                        fill=#1!30
                    },
                arrow/.style={->, thick},
                dashedarrow/.style={->, thick, dashed}
            ]

            \node[box=PastelPurple] (eis)
            {measured\\EIS $(Z', Z'')$};

            \node[box=PastelGreen, right=1.5cm of eis] (cnn)
            {\gls{cnn}\\outputs $g(\log\tau)$};

            \node[box=PastelYellow, right=1.7cm of cnn] (recon)
            {numerical solution of\\[2pt]
                \scriptsize
                $Z(\omega)=R_s+\!\int g(\log\tau)\frac{1-j\omega\tau}{1+(\omega\tau)^2}d\log\tau + j\omega L$};

            \node[box=PastelPink, right=1.7cm of recon] (zrec)
            {reconstructed\\impedance $\hat Z$};

            \node[box=PastelRed, below=2cm of recon] (loss)
            {loss:\\$\hat Z$ vs.\ $Z$};

            \node[smallbox=PastelGreen, above=1.1cm of recon] (rs)
            {$R_s$ network\\{\scriptsize predicts $R_s$ from $Z'$}};

            \node[smallbox=PastelOrange,
                below=1cm of cnn,
                xshift=2.5cm] (lpar)
            {trainable\\parameter $L$};

            \node[
                draw,
                rounded corners,
                fill=PastelBlue!40,
                minimum width=4.4cm,
                minimum height=3.4cm,
                below=0.8cm of loss,
            ] (latent)
            {
                \begin{minipage}{4.7cm}
                    \centering
                    \textbf{latent space analysis}\\[-2pt]
                    {\scriptsize (post training)}

                    \vspace{3pt}
                    \includegraphics[width=4.5cm]{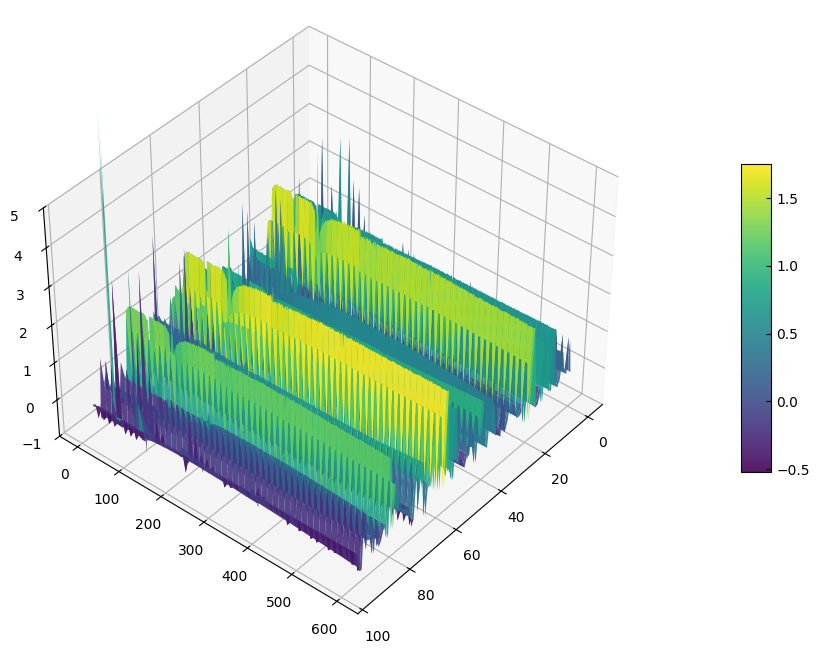}
                \end{minipage}
            };

            \draw[arrow] (eis) -- node[above]{\scriptsize 2 channels} (cnn);
            \draw[arrow] (cnn) -- node[above]{\scriptsize $g(\log\tau)$} (recon);
            \draw[arrow] (recon) -- (zrec);

            \draw[arrow] (eis.north) |- (rs.west);
            \draw[arrow] (rs.south) -- node[right]{\scriptsize $R_s$} (recon.north);

            \draw[arrow] (lpar.east) -| node[pos=0.72, right]{\scriptsize $j\omega L$} (recon.south);

            \draw[arrow] (zrec.south) |- (loss.east);
            \draw[arrow] (loss.west) -| (cnn.south);

            \draw[dashedarrow] ($(cnn.south)+(-0.35,0)$) |- (latent.west);

        \end{tikzpicture}%
    }
    \caption{Schematic representation of the proposed physics-informed \gls{cnn} framework.
        Measured impedance spectra are provided as two input channels and mapped by the \gls{cnn} to the corresponding \gls{drt}, $g(\log\tau)$.
        In parallel, an auxiliary branch estimates the series resistance $R_s$ from the real part of the impedance, while the inductive contribution $L$ is represented by a trainable parameter.
        The predicted \gls{drt}, $R_s$, and $L$ are inserted into the physics-informed reconstruction layer to obtain the reconstructed impedance $\hat Z$.
        The model is trained by comparing $\hat Z$ with the measured impedance $Z$, enabling physics-informed learning through impedance reconstruction error.
        After training, the learned latent representation is used for interpretability and condition monitoring analyses.
    }\label{fig:pipeline}
\end{figure}
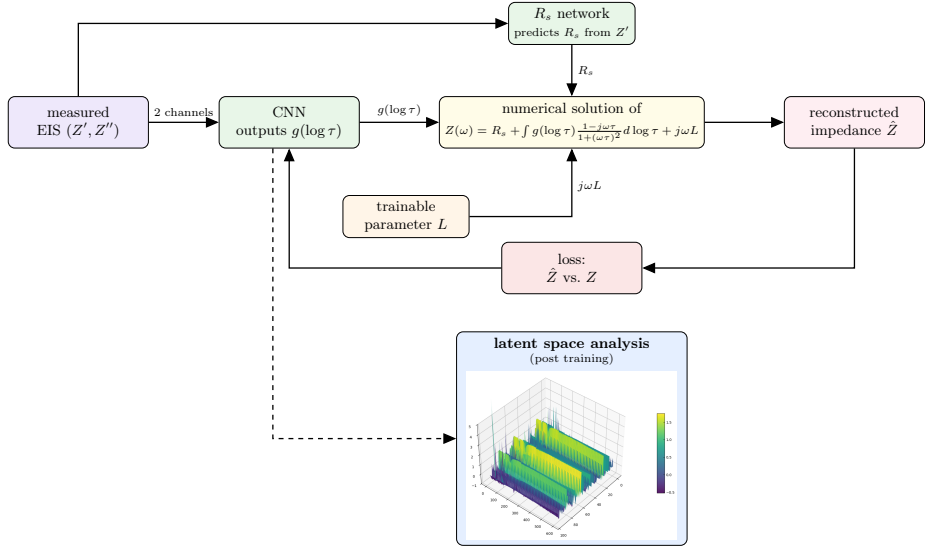

The complete training procedure is illustrated in \figurename~\ref{fig:pipeline}.
The measured real and imaginary components of the impedance spectrum are provided as input to the \gls{cnn}, which predicts the corresponding \gls{drt}.
In addition to the \gls{drt}, a separate neural network predicts the series resistance \(R_s\), while the inductive contribution \(L\) is represented by a trainable parameter.
The predicted distribution is then inserted into the discretised electrochemical model and transformed back into the impedance domain using the trapezoidal approximation.
The reconstructed impedance is compared with the measured spectrum to compute the training loss,
\begin{equation}
    \mathcal{L}
    =
    \mathrm{MSE}
    \!\left(
    \hat{Z}^{\prime},
    Z^{\prime}
    \right)
    +
    \mathrm{MSE}
    \!\left(
    \hat{Z}^{\prime\prime},
    Z^{\prime\prime}
    \right)
    +
    \lambda_L L^2 ,
\end{equation}
where primes denote the real components, double primes denote the imaginary components, and \(\lambda_L = 10^6\) penalises large inductive contributions.

Because the impedance reconstruction layer consists entirely of differentiable operations, gradients can be propagated through the numerical integration procedure during backpropagation.
Consequently, the parameters of the \gls{cnn}, the resistance model \(R_s\), and the inductive parameter \(L\) are jointly optimised using the Adam optimiser.

As a result, the network is trained to produce \glspl{drt} that are consistent with the underlying electrochemical relationship rather than merely fitting the impedance spectrum.
After training, the encoder additionally provides a compact latent representation of the impedance data.
This latent space can be analysed independently of the reconstruction task and is used in the subsequent sections to investigate latent space organisation, condition monitoring, and long term system evolution.

By embedding the governing electrochemical relation directly into the optimisation procedure, the network is constrained to produce physically consistent \glspl{drt} while retaining the flexibility of a data driven model.
\FloatBarrier
\subsection{Mitigating checkerboard artifacts}
\begin{figure}[h]
    \centering
    \includegraphics{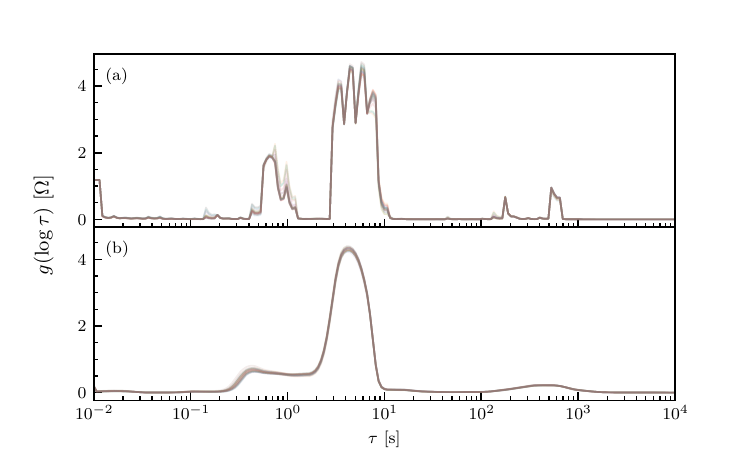}
    \caption{Comparison of decoder architectures.\ (a) Reconstructions obtained using transposed convolutions exhibit characteristic checkerboard artifacts.\ (b) Replacing transposed convolutions with an upsampling layer followed by a standard convolution removes these artifacts and produces smoother \gls{drt} estimates.}\label{fig:checkerboard}
\end{figure}

Transposed convolution layers can produce checkerboard artifacts when kernel size, stride, and padding lead to uneven overlap during upsampling~\citep{odena2016deconvolution}.
In \gls{drt} estimation, such artifacts manifest as spurious oscillations that may be misinterpreted as physical relaxation processes (\figurename~\ref{fig:checkerboard}a).
To eliminate this effect, each transposed convolution was replaced by an upsampling layer followed by a standard convolution, producing smooth reconstructions without artificial structures (\figurename~\ref{fig:checkerboard}b).
The checkerboard artifacts are effectively removed while preserving the overall shape and locations of the reconstructed relaxation processes.

Beyond improving reconstruction quality, this modification promotes smoother and more physically plausible \gls{drt} estimates, which is important for the reliable interpretation of the resulting distributions.

\subsection{Architecture selection}\label{sec:architecture_study}

In solving the inverse \gls{drt} problem, the primary challenge is not merely minimizing the reconstruction error but preventing the amplification of measurement noise.
Models with excessive capacity tend to fit stochastic fluctuations in the impedance spectra, producing non-physical oscillations in the reconstructed \glspl{drt}.
Conversely, models with insufficient capacity fail to resolve overlapping relaxation processes.

To determine an appropriate balance between reconstruction fidelity and model complexity, a grid search across 216 network architectures was performed.
The search space included variations in network depth, channel width, bottleneck size, pooling strategy, and receptive field.

\begin{figure}[h]
    \centering
    \includegraphics{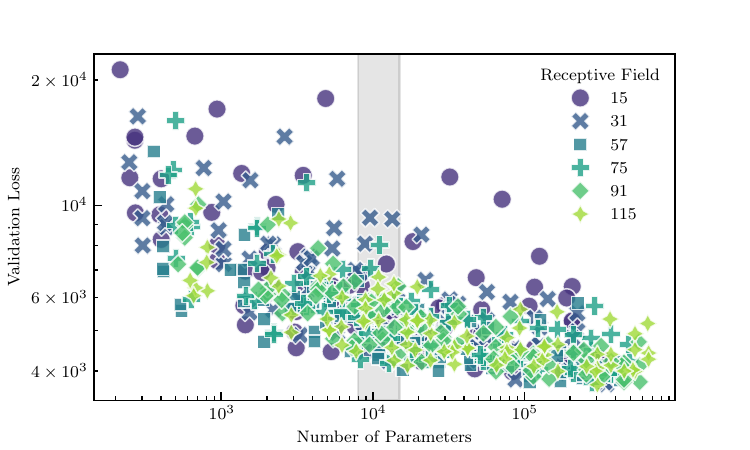}
    \caption{Validation loss as a function of model complexity.}\label{fig:l_curve}
\end{figure}

The relationship between model complexity and validation loss is shown in \figurename~\ref{fig:l_curve}.
A distinct elbow is observed at approximately $10^4$ trainable parameters.
Below this threshold, the model underfits the impedance spectra and exhibits elevated reconstruction error.
Beyond approximately $1.5\times10^4$ parameters, improvements become marginal, indicating diminishing returns and an increased risk of overfitting.

A more detailed analysis of the architectural search is provided in
Section~S1 of the Supplementary Material.
Based on the complete study, a receptive field of 57 and a bottleneck size of 160 were selected as the best compromise between reconstruction accuracy, stability, and resistance to overfitting.
The resulting architecture contains approximately 12,000 trainable parameters and was used throughout the remainder of this work.

Importantly, the selected architecture was not optimized for a specific experiment.
Instead, the same network configuration was subsequently applied to all investigated datasets without any dataset-specific redesign or hyperparameter tuning.
This provides a stringent test of the framework's ability to generalize across different operating conditions, degradation mechanisms, and experimental campaigns.
\FloatBarrier
\subsection{Experimental datasets}\label{sec:datasets}

To evaluate the generalisation capability of the proposed framework, the selected architecture was applied to three independent experimental datasets covering different operating regimes, degradation mechanisms, and experimental objectives.

\textbf{\Gls{sofc} stack monitoring campaign.}
The first dataset consists of 603 \gls{eis} measurements acquired during a
3600-hour operation of a six-cell \gls{sofc} stack, including several
fuel starvation events, emergency shutdowns due to H$_2$ shortage, and an
unplanned power loss~\cite{NUSEV2021,Boskoski2024,Gradisar2026}.
This dataset was used to evaluate impedance reconstruction accuracy, latent space condition monitoring, and event detection.

\textbf{Operating condition study.}
The second dataset corresponds to the experimental study presented
in~\cite{boskoski2024extracting}, where an industrial-scale \gls{soec}
(100~cm$^{2}$) was operated under three representative conditions: high hydrogen content at moderate current density (Condition~1, 700~mA~cm$^{-2}$, 20\%~H$_2$, 60~h), low hydrogen content at the same current density (Condition~2, 700~mA~cm$^{-2}$, 10\%~H$_2$, 85~h), and elevated current density with unstable steam supply (Condition~3, 900~mA~cm$^{-2}$, 120~h).
Changes were originally characterised using frequency-resolved \gls{kl} divergence analysis, and here the dataset is used to examine whether the learned latent representation captures the same operating-state transitions.

\textbf{Multi-regime degradation campaign.}
The third dataset consists of 1248 \gls{eis} spectra collected during a
2650-hour experimental campaign on a commercial $4{\times}4$~cm$^{2}$
electrolyte-supported cell~\cite{MUTTER2026239640}.
The campaign comprised six sequential degradation phases: steam electrolysis at
300~mA~cm$^{-2}$ (P1, 1000~h), co-electrolysis at 300~mA~cm$^{-2}$ (P2, 400~h),
and 500~mA~cm$^{-2}$ (P3, 200~h), a return to steam electrolysis (P4, 300~h),
reversible EC/FC operation at 300/$-$150~mA~cm$^{-2}$ (P5, 264~h), and a final
steam electrolysis phase (P6, 300~h).
This dataset was used to investigate long term \gls{drt} evolution and to assess whether latent space trajectories capture transitions between different operating regimes and degradation processes.




\section{Results and discussion}
The results are presented in four stages. 
First, the framework is evaluated on experimental data to verify impedance reconstruction accuracy, physics consistency, and comparison with existing \gls{drt} estimation methods. 
Second, the proposed framework is validated on synthetic spectra with analytically known distributions to assess its ability to resolve overlapping relaxation processes. 
Third, the learned latent representation is analysed to investigate its physical organisation and suitability for condition monitoring. 
Finally, the framework is applied to independent experimental datasets to assess the generalisability of the complete analysis pipeline.
\subsection{Core validation and physics consistency}\label{sec:phase1}

The primary objective of the proposed physics-informed framework is to estimate \glspl{drt} that remain consistent with the governing electrochemical relation in~\eqref{eq:drt_original}.
Since the predicted \glspl{drt} are transformed back into the impedance domain through the embedded physical model, accurate reconstruction of the measured impedance spectra provides direct evidence that the network satisfies the underlying electrochemical constraints.

\begin{figure}[h]
    \centering
    \includegraphics[width=\linewidth]{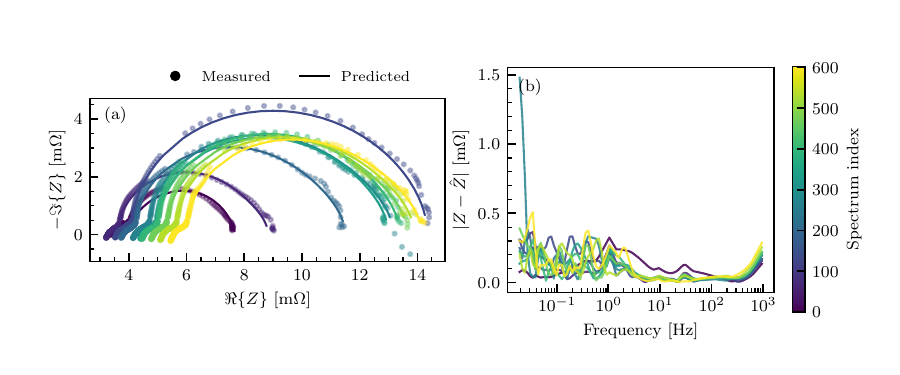}
    \caption{Representative reconstruction results for the \gls{sofc} stack
        monitoring dataset.~(a) Measured and reconstructed Nyquist spectra for selected
        measurements.~(b) Frequency-dependent reconstruction error, $|Z-\hat{Z}|$.
        The low residuals confirm accurate impedance reconstruction from the predicted
        \glspl{drt}.}\label{fig:processed_eis}
\end{figure}

Representative reconstruction results obtained for the \gls{sofc} stack monitoring dataset introduced in Section~\ref{sec:datasets} are shown in \figurename~\ref{fig:processed_eis}.
The reconstructed Nyquist spectra closely follow the measured impedance response, while the residual errors remain small across the investigated frequency range.

The corresponding Nyquist and residual plots for the operating condition study
and the multi-regime degradation campaign are provided in Section~S2 of the
Supplementary Material.

Reconstruction accuracy across all three datasets was quantified using the \gls{rmse} between measured and reconstructed impedance spectra, complemented by a range-normalised \gls{rmse} to enable comparison across datasets with different impedance magnitudes.

\begin{table}[h]
    \centering
    \resizebox{\linewidth}{!}{%
        \begin{tabular}{@{}lcccc@{}}
            \toprule
            \textbf{Dataset}            &
            \textbf{RMSE (Re)}          &
            \textbf{RMSE (Im)}          &
            \textbf{RMSE (total)}       &
            \textbf{R-RMSE}                                             \\
                                        &
            \textbf{[m$\Omega$]}        &
            \textbf{[m$\Omega$]}        &
            \textbf{[m$\Omega$]}        &
            \textbf{[\%]}                                               \\
            \midrule
            \gls{sofc} stack monitoring & 0.536 & 0.191 & 0.402 & 0.291 \\
            Operating condition study   & 1.461 & 1.741 & 1.607 & 1.062 \\
            Multi-regime degradation    & 0.504 & 0.453 & 0.479 & 0.451 \\
            \bottomrule
        \end{tabular}
    }
    \caption{Impedance reconstruction accuracy for the investigated datasets.
        RMSE values are reported together with the range-normalised RMSE (R-RMSE),
        expressed relative to the impedance range of each dataset.}\label{tab:reconstruction_metrics}
\end{table}

Despite substantial differences in operating conditions, degradation mechanisms, and experiment duration, the reconstruction errors remain consistently low across all datasets.
In all cases, the range-normalised \gls{rmse} remains below approximately $1.1\%$, demonstrating excellent agreement between the measured and reconstructed impedance spectra.
These results provide empirical validation of the physics-informed nature of the proposed architecture.

To further assess the quality of the estimated \glspl{drt}, the proposed approach was compared with conventional Tikhonov regularisation and a previously reported neural network-based method.

\begin{figure}[h]
    \centering
    \includegraphics[width=\linewidth]{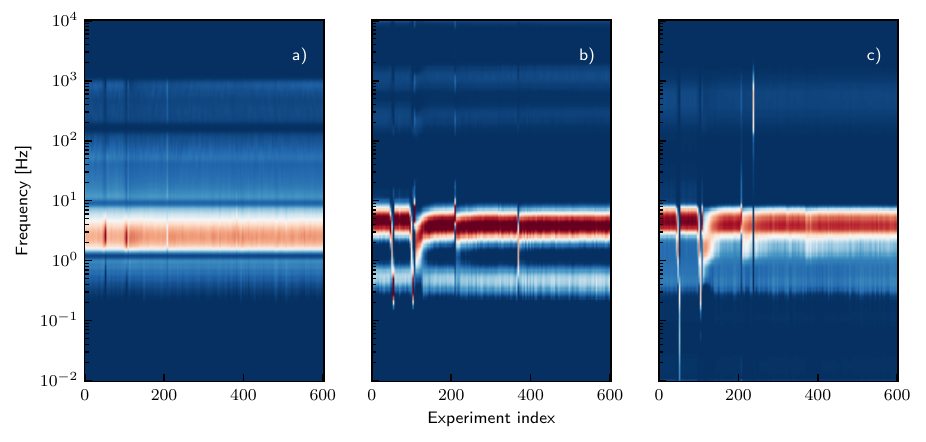}
    \caption{Evolution of \glspl{drt} over time using (a) Tikhonov
        regularisation~\cite{10.1016/j.electacta.2015.09.097}, (b) a feed-forward
        neural network-based
        approach~\cite{10.1149/1945-7111/ab631a,Gradisar2026},
        and (c) the proposed \gls{cnn} framework.}\label{fig:compared_drt_results}
\end{figure}

\figurename~\ref{fig:compared_drt_results} compares \gls{drt} estimates obtained using Tikhonov regularisation~\cite{tikhonov}, a feed-forward neural network approach~\cite{10.1149/1945-7111/ab631a}, and the proposed \gls{cnn} framework.
All three methods identify the dominant relaxation process, visible as the pronounced band between approximately 1 and 10~Hz, but differ substantially in how they represent its temporal evolution.
Tikhonov regularisation (\figurename~\ref{fig:compared_drt_results}a) produces a smooth, nearly stationary distribution, suppressing much of the temporal variation observed throughout the experiment.
The feed-forward neural network (\figurename~\ref{fig:compared_drt_results}b) recovers considerably richer structure, including the gradual emergence of a secondary lower-frequency relaxation process, but at the cost of several abrupt transitions between neighbouring experiment indices, most notably early in the experiment and again around index 200.
The proposed \gls{cnn} framework (\figurename~\ref{fig:compared_drt_results}c) captures the same temporal features as the feed-forward approach while evolving more continuously between successive measurements.
Abrupt transitions still appear at a few experiment indices, but they are markedly less pronounced.

This continuity follows directly from training a single shared model across the entire dataset.
Rather than fitting each spectrum with an independently chosen regularisation strength, as Tikhonov regularisation and per-spectrum neural networks both do, the proposed framework applies the same implicit regularisation to every measurement, allowing neighbouring spectra to be compared directly rather than through the lens of separately tuned fits.
The value of this consistency lies not in superior point-wise accuracy, since all three methods agree closely on the dominant relaxation process, but in ensuring that differences observed across the dataset reflect genuine electrochemical changes rather than fitting artefacts.
This is confirmed quantitatively by the sub-$1.1\%$ reconstruction error achieved throughout, showing that the network has learned the governing electrochemical relationship rather than a statistical shortcut.

\subsection{Validation on synthetic overlapping relaxation processes}\label{sec:zarc_validation}

The ability to distinguish closely spaced relaxation processes was further assessed using synthetic two-ZARC spectra with analytically known \glspl{drt}.
A ZARC element, consisting of a resistor in parallel with a constant-phase element, is described by
\begin{equation}
    Z_{\mathrm{ZARC}}(\omega)
    =
    \frac{R}{1+(\mathrm{j}\omega)^{\phi}RQ}.
    \label{eq:zarc}
\end{equation}
The total impedance is given by
$Z(\omega)=R_s+Z_{\mathrm{ZARC},1}(\omega)+Z_{\mathrm{ZARC},2}(\omega)$,
where $R$ is the process resistance, $Q$ is the constant-phase-element parameter, $\phi$ is the dispersion parameter, and $R_{\mathrm{s}}$ is the series resistance.
For a known impedance transfer function, the corresponding analytical \gls{drt} was obtained by analytic continuation using the Fuoss--Kirkwood formulation~\cite{fuoss1941electrical},
\begin{equation}
    \begin{aligned}
        G(u)
         & =
        -\frac{1}{\pi}
        \left[
            \Im\!\left\{
            Z\!\left(e^{-u-\mathrm{j}\pi/2}\right)
            \right\}
            +
            \Im\!\left\{
            Z\!\left(e^{-u+\mathrm{j}\pi/2}\right)
            \right\}
            \right],
    \end{aligned}
    \label{eq:fuoss_kirkwood_drt}
\end{equation}
where $u=\log\tau$.
This analytical solution provides an exact reference against which the recovered peak structure and relaxation-time positions can be compared.
To evaluate the proposed framework, the architecture selected in Section~\ref{sec:architecture_study} was trained on 1,000 synthetic two-ZARC spectra generated by varying the resistances, dispersion parameters $\phi$, and the logarithmic separation between the characteristic relaxation times.
Peak separations spanning 0.2--2.5 decades were included to cover both strongly overlapping and well-separated relaxation processes.
The complete parameter ranges and data generation procedure are provided in Appendix~\ref{app:synthetic_zarc}.
The validation spectra presented below were generated separately and were not included in the training dataset.

Following the validation protocol of~\citet{10.1149/1945-7111/ab631a}, two representative test cases were considered, with relaxation times separated by two decades and one decade, respectively.
To evaluate the network under non-ideal conditions, independent Gaussian noise with a standard deviation of $\sigma=0.5~\Omega$ was added to both the real and imaginary impedance components.
This corresponds to the upper limit of the noise range used during training and therefore represents a more demanding validation than the noise-free spectra considered in the original study.

\begin{figure}[h]
    \centering
    \includegraphics[width=\linewidth]{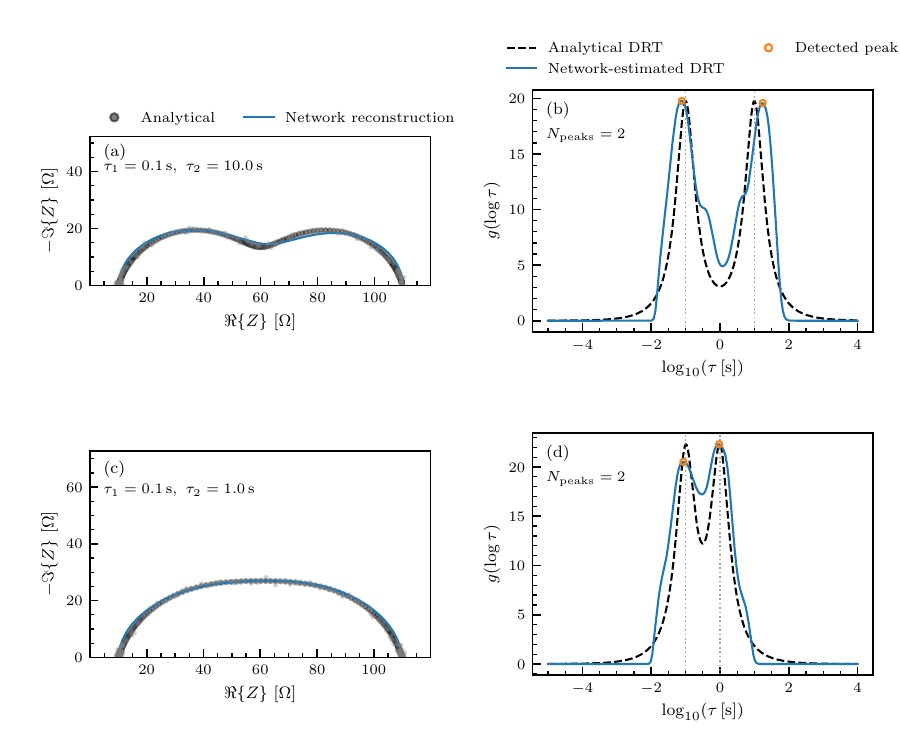}
    \caption{Validation on synthetic two-ZARC spectra with additive Gaussian noise ($\sigma=0.5~\Omega$), using the test configurations considered by~\citet{10.1149/1945-7111/ab631a}. (a)--(b) Two-decade separation with $\tau_1=0.1$~s and $\tau_2=10$~s. (c)--(d) One-decade separation with $\tau_1=0.1$~s and $\tau_2=1$~s. The left panels compare the analytical impedance spectra with the network reconstructions, while the right panels compare the analytical and network-estimated \glspl{drt}.}
    \label{fig:zarc_validation}
\end{figure}

The results are shown in \figurename~\ref{fig:zarc_validation}.
For the two-decade separation, the two relaxation processes are already reflected by partially separated arcs in the Nyquist spectrum (\figurename~\ref{fig:zarc_validation}a).
The corresponding network-estimated \gls{drt} accurately reproduces the analytical two-peak structure, with both relaxation processes clearly identified (\figurename~\ref{fig:zarc_validation}b).
Although the higher-$\tau$ peak is shifted by 0.241 decades, the two processes remain clearly resolved.

The one-decade case provides a considerably more challenging example.
Here, the two ZARC contributions overlap strongly and appear as a single broad depressed arc in the Nyquist representation (\figurename~\ref{fig:zarc_validation}c), such that the presence of two underlying relaxation processes is not readily apparent from visual inspection of the impedance spectrum.
Nevertheless, the network-estimated \gls{drt} reproduces the analytical two-peak structure with good agreement (\figurename~\ref{fig:zarc_validation}d), with peak-position errors of only 0.065 and 0.025 decades.

These results demonstrate that the proposed framework accurately recovers the analytical \gls{drt} for overlapping two-ZARC spectra.
In the one-decade case, the network reproduces the correct two-peak structure even though the corresponding Nyquist spectrum appears as a single broad depressed arc.
This indicates that the learned impedance-to-\gls{drt} mapping exploits subtle information contained across the full impedance spectrum rather than relying solely on features that are visually apparent in the Nyquist representation.

\FloatBarrier
\subsection{Interpretable latent space dynamics}\label{sec:phase2}

The question is whether the learned latent representation goes further by containing interpretable information about the electrochemical system beyond what is needed for reconstruction.
In particular, we examine whether the latent space exhibits a meaningful internal structure and whether its evolution can be used to monitor changes in system operation.

At time $t$, the encoder maps the input impedance spectrum to a latent tensor
$\mathbf{Z}_t\in\mathbb{R}^{C\times P}$, where $C=40$ denotes the number of output channels produced by the final convolutional layer and $P=4$ the pooled positions retained after adaptive pooling.
Each channel represents one learned component of the encoded impedance spectrum, while the pooled positions preserve coarse information about where that component originated along the frequency axis before pooling.
Flattening $\mathbf{Z}_t$ row-wise yields the latent vector $\mathbf{z}_t\in\mathbb{R}^{CP}$ used throughout the remainder of this section.

\begin{figure}[h]
    \centering
    \includegraphics[width=\linewidth]{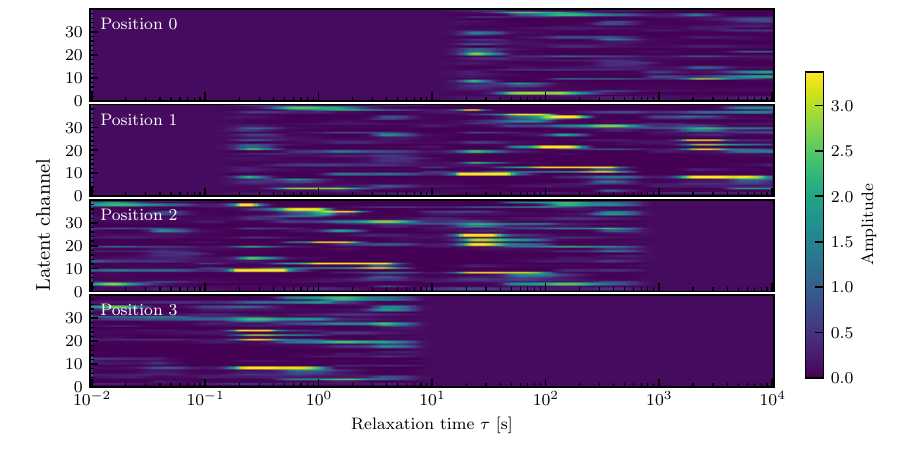}
    \caption{Decoder probe analysis of the learned latent basis. For each pooled position, every latent channel was activated individually while all remaining latent entries were fixed to zero. Each row corresponds to one latent channel, ordered according to the location of its peak response at pooled position~0. This ordering is preserved across all panels. Colour indicates the amplitude of the reconstructed relaxation-time response produced by activating the corresponding channel--position pair.}\label{fig:latent_probe}
\end{figure}

To investigate the physical meaning of these latent components, a decoder probe analysis was performed.
For every channel--position pair $(c,p)$, an artificial latent tensor $\tilde{\mathbf{Z}}\in\mathbb{R}^{C\times P}$ was constructed with all entries set to zero except $\tilde{Z}_{c,p}$, which was assigned a fixed activation.
Passing this tensor through the decoder reveals the relaxation-time response associated with activating only that individual latent component.
Repeating this procedure for all $40\times4$ channel--position pairs produces the activation maps shown in \figurename~\ref{fig:latent_probe}, where each panel corresponds to one pooled position and each row to one latent channel.
The channels are ordered according to the location of their peak response at pooled position~0, and this ordering is retained across all panels to facilitate comparison.

The decoder associates the four pooled positions with progressively different regions of the relaxation-time axis.
For each channel, the peak of the reconstructed response was identified, and the median peak location across all channels was computed for each pooled position (the median was used because relaxation times span several decades).
Pooled positions~0 and~1 predominantly activate long-timescale processes, with median peak responses at approximately $\tau=24$~s and $\tau=83$~s, respectively, whereas positions~2 and~3 shift towards shorter timescales, with median peak responses near $\tau=3.5$~s and $\tau=0.3$~s.

Within each pooled position, different channels respond to different parts of the corresponding relaxation-time range, providing a finer subdivision of the represented electrochemical processes.
The latent representation therefore exhibits a hierarchical organisation: the pooled position determines a coarse relaxation-time region, while the channel identity provides additional resolution within that region.
Importantly, this organisation was not imposed during training.
The network was optimised solely to reconstruct the measured impedance, without any supervision encouraging latent components to specialise by relaxation timescale.
The emergence of this structure indicates that the learned latent space reflects the physical organisation of the underlying electrochemical processes rather than constituting an arbitrary compressed representation.

This physically organised latent representation also provides a natural space for quantifying changes in system behaviour over time.
Rather than comparing impedance spectra directly, system evolution can be monitored by measuring distances between the corresponding latent representations.
The Euclidean distance between the latent vector at time $t$ and a chosen reference vector $\mathbf{z}_{\mathrm{ref}}$ is defined as

\begin{equation}
    D_{\mathrm{ref}}(t)
    =
    \left\lVert
    \mathbf{z}_t-\mathbf{z}_{\mathrm{ref}}
    \right\rVert_2 .
    \label{eq:latent_distance_general}
\end{equation}

To demonstrate this, the \gls{sofc} stack monitoring dataset was analysed using $D_{10}(t)$, where the 10th measurement serves as the nominal reference state.

\begin{figure}[h]
    \centering
    \includegraphics[width=\linewidth]{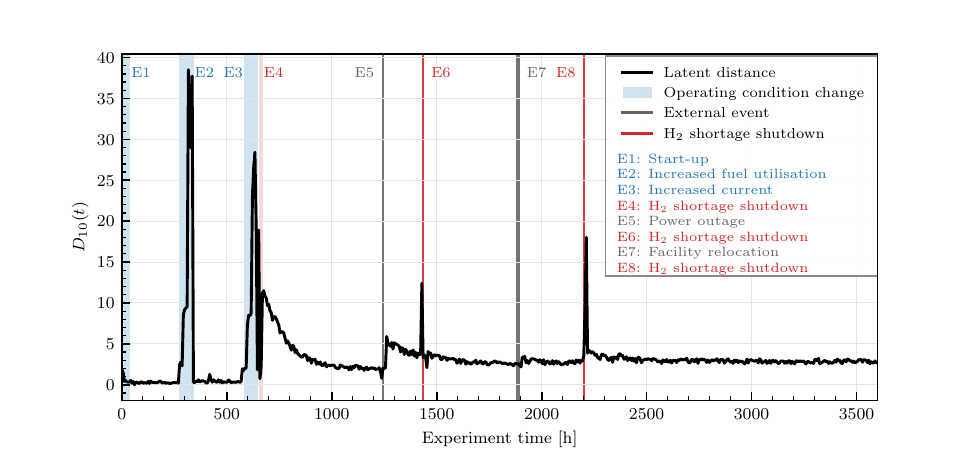}
    \caption{Latent space condition monitoring for the \gls{sofc} stack monitoring dataset. $D_{10}(t)$ is used as an indicator of system evolution. Pronounced peaks correspond to major operational events during the campaign.}\label{fig:condition_monitoring}
\end{figure}

The resulting distance profile exhibits pronounced peaks that coincide with known operational events during the experimental campaign, summarised in \tablename~\ref{tab:cea}.

\begin{table}[h]
    \centering
    \resizebox{\linewidth}{!}{%
        \begin{tabular}{@{}lll@{}}
            \toprule
            \textbf{Event number} & \textbf{Experiment time [h]} & \textbf{Description}                                  \\
            \midrule
            E1                    & 0--40                        & Start up                                              \\
            E2                    & 270--342                     & Increased fuel utilisation (decreased fuel flow rate) \\
            E3                    & 582--648                     & Increased fuel utilisation (increased current)        \\
            E4                    & 654--672                     & Emergency shutdown due to H$_2$ shortage              \\
            E5                    & 1242                         & Power loss due to thunderstorms                       \\
            E6                    & 1434                         & Emergency shutdown due to H$_2$ shortage              \\
            E7                    & 1884--1890                   & Moving into the new building                          \\
            E8                    & 2202                         & Emergency shutdown due to H$_2$ shortage              \\
            \bottomrule
        \end{tabular}
    }
    \caption{Operational events corresponding to the peaks observed in
        \figurename~\ref{fig:condition_monitoring}~\cite[same as Table~1]{NUSEV2021}.}\label{tab:cea}
\end{table}

Increases in $D_{\mathrm{10}}(t)$, seen as peaks in \figurename~\ref{fig:condition_monitoring}, coincide with fuel starvation events, shutdowns, and other major perturbations, demonstrating that deviations from the nominal operating state are reflected directly and immediately in the latent representation.
These observations are consistent with the conclusions of~\cite{Gradisar2026}, where significant events were detected through comparisons of reconstructed \glspl{drt} using \gls{uot}.
While \gls{uot} provides a principled metric for comparing \glspl{drt}, the latent space distance~\eqref{eq:latent_distance_general} offers a computationally efficient alternative obtained directly from the encoder output, avoiding ambiguous \gls{drt} curve comparisons.

The decoder probe analysis and the condition monitoring results confirm that the learned latent representation is both physically interpretable and operationally informative.
The network not only reconstructs impedance spectra accurately, but also spontaneously organises electrochemical information into a structured latent space, a physically interpretable representation that was never explicitly supervised and yet reliably identifies degradation and operational events.

\FloatBarrier
\subsection{Pipeline generalisation}\label{sec:phase3}

The previous sections established that the proposed framework produces physically consistent \gls{drt} estimates and learns an interpretable latent representation from a single experimental campaign.
The critical remaining question is whether the same pipeline can extract equally meaningful information from fundamentally different experiments, without any modification to the architecture, hyperparameters, or analysis workflow.

To answer this, the framework was applied to two independent datasets that differ from the \gls{sofc} stack monitoring campaign in cell type, operating regime, degradation mechanism, and experimental objective.
The results therefore constitute a direct test of architectural generalisation, as the same 12{,}000-parameter network, trained independently on each dataset, required no redesign, retuning, or workflow modification.

\paragraph{Operating condition study~\cite{boskoski2024extracting}}

\begin{figure}[h]
    \centering
    \includegraphics[width=\linewidth]{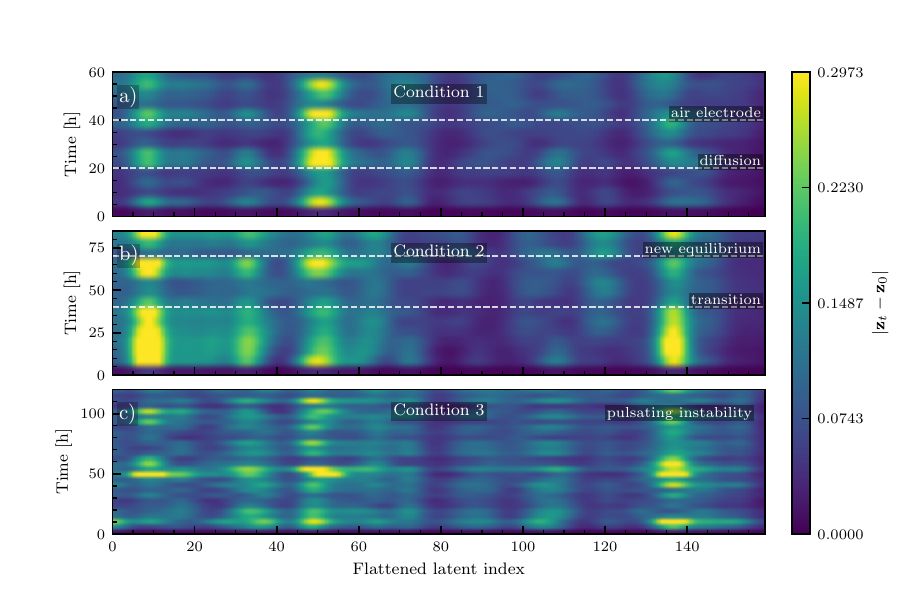}
    \caption{Latent space evolution for the three operating conditions investigated
        in~\cite{boskoski2024extracting}. Heatmaps show the smoothed absolute difference between each latent representation and the initial spectrum,
        $\left|\mathbf{z}_t - \mathbf{z}_0\right|$, where brighter regions indicate larger cumulative deviation from the initial operating state. Dashed lines denote characteristic time points reported in the original analysis.}\label{fig:conditions123_stacked}
\end{figure}

This dataset (Section~\ref{sec:datasets}) comprised three operating conditions investigated over 60--125~h each on an industrial-scale \gls{soec}.
In the original study, changes in system behaviour were identified using \gls{kl} divergence computed frequency by frequency relative to an initial reference spectrum, identifying frequency bands where the impedance changed statistically significantly~\cite{boskoski2024extracting}.

Here, the same intuition is applied in latent space.
Recall from Section~\ref{sec:phase2} that each spectrum's latent tensor is flattened into a vector $\mathbf{z}_t \in \mathbb{R}^{CP}$.
The scalar distance $D_{\mathrm{ref}}(t)$ from~\eqref{eq:latent_distance_general} summarises the overall change between $\mathbf{z}_t$ and a reference $\mathbf{z}_{\mathrm{ref}}$ as a single number, but does not indicate which part of the latent representation changed.
To recover this information, we instead compute the element-wise absolute difference $\left|\mathbf{z}_t - \mathbf{z}_0\right|$: rather than combining all $CP$ entries into one value, this keeps a separate difference for each entry, so that changes can be localised to specific channels and pooled positions, in the same way that the original study localises changes to specific frequency bands.
This correspondence is temporal rather than spectral: the original study attributes each transition to a specific frequency band, whereas here we only identify the same transition times through changes in latent activation, without establishing which latent channels or positions correspond to which frequency band.
Unlike the original approach, this requires neither frequency-band selection nor numerical integration over the impedance spectrum, since the diagnostic information is extracted directly from the encoder output as a byproduct, with no additional analysis required.
The resulting activation maps are shown in \figurename~\ref{fig:conditions123_stacked}.

For Condition~1, diffusion-related losses emerging after approximately 20~h (0.3--2~Hz) and air-electrode changes after approximately 40~h (10--40~Hz), originally identified through \gls{kl} divergence~\cite{boskoski2024extracting}, coincide in time with changes in latent activation visible in \figurename~\ref{fig:conditions123_stacked}a.
The same holds for Condition~2, where the gradual convergence toward a new equilibrium, marked by spectral changes shifting from 200--800~Hz near 40~h to 800~Hz--4~kHz near 70~h~\cite{boskoski2024extracting}, coincides with two distinct transitions in \figurename~\ref{fig:conditions123_stacked}b.
Condition~3 differs in character: the persistent pulsating instability caused by inconsistent steam supply at 900~mA~cm$^{-2}$~\cite{boskoski2024extracting} does not produce discrete transition events.
Instead, \figurename~\ref{fig:conditions123_stacked}c shows sustained, recurring latent activity throughout the experiment, mirroring the continuous operational instability rather than isolated degradation onsets.

\paragraph{Multi-regime degradation campaign~\cite{MUTTER2026239640}}

\begin{figure}[h]
    \centering
    \includegraphics[width=\linewidth]{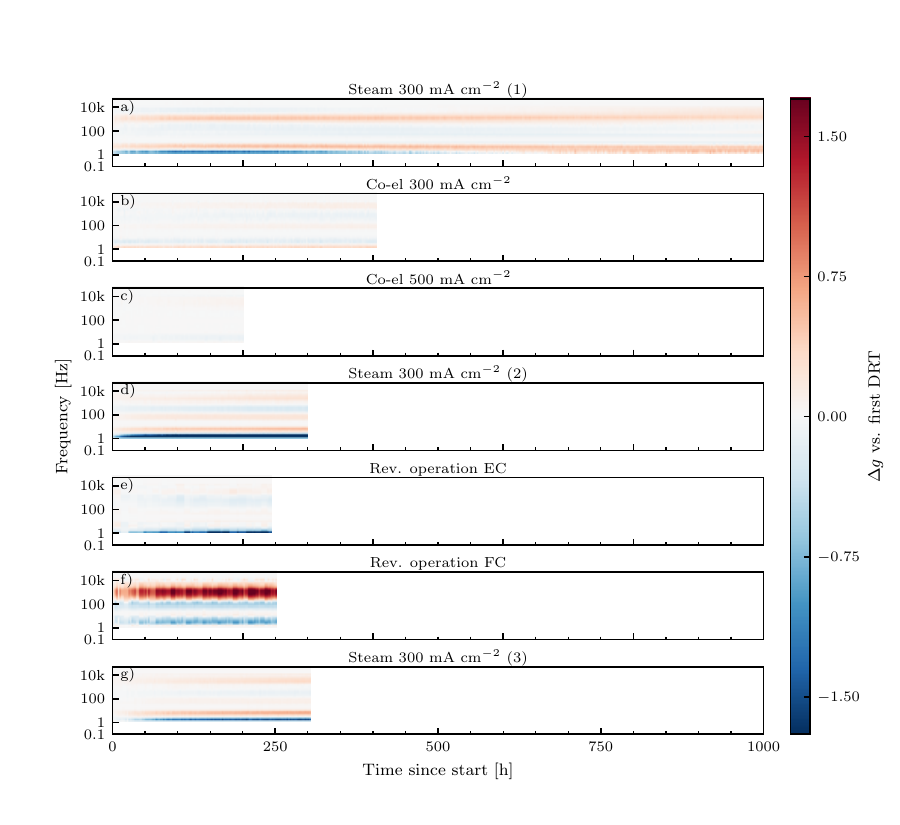}
    \caption{Difference between the network-estimated \glspl{drt} and the first
        estimated \gls{drt} within each operating regime
        ($\Delta g_t(\log\tau)$,~\eqref{eq:drt_diff}). Horizontal axis shows elapsed
        experiment time within each phase.}\label{fig:delta_drt}
\end{figure}

\begin{figure}[h]
    \centering
    \includegraphics[width=\linewidth]{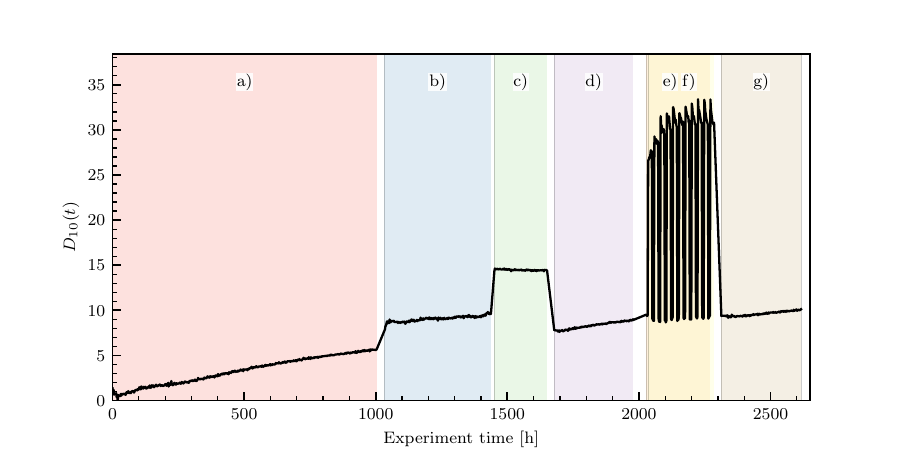}
    \caption{Evolution of $D_{10}(t)$ during the 2650-hour campaign.
        Coloured regions denote the operating regimes in~\cite{MUTTER2026239640}.}\label{fig:graz_condition_monitoring}
\end{figure}

This dataset (Section~\ref{sec:datasets}) comprised 1248 \gls{eis} spectra collected on a commercial $4{\times}4$~cm$^{2}$ electrolyte-supported cell, with degradation characterised phase by phase through bi-hourly \gls{drt} diagnostics~\cite{MUTTER2026239640}.
\tablename~\ref{tab:graz_phases} summarises the six operating phases and their correspondence to the panels in \figurename~\ref{fig:delta_drt} and \figurename~\ref{fig:graz_condition_monitoring}.

\begin{table}[h]
    \centering
    \resizebox{\linewidth}{!}{%
        \begin{tabular}{@{}llll@{}}
            \toprule
            \textbf{Phase}               &
            \textbf{Panels}              &
            \textbf{Experiment time [h]} &
            \textbf{Operating regime}                                                                       \\
            \midrule
            P1                           & a)     & 22--1022   & Steam electrolysis, 300~mA~cm$^{-2}$       \\
            P2                           & b)     & 1055--1455 & Co-electrolysis, 300~mA~cm$^{-2}$          \\
            P3                           & c)     & 1473--1673 & Co-electrolysis, 500~mA~cm$^{-2}$          \\
            P4                           & d)     & 1701--2001 & Steam electrolysis, 300~mA~cm$^{-2}$       \\
            P5                           & e)--f) & 2048--2312 & Reversible EC/FC operation                 \\
            P6                           & g)     & 2335--2635 & Final steam electrolysis, 300~mA~cm$^{-2}$ \\
            \bottomrule
        \end{tabular}
    }
    \caption{Operating phases of the degradation campaign investigated in~\cite{MUTTER2026239640} and their correspondence to the panels in \figurename~\ref{fig:delta_drt} and \figurename~\ref{fig:graz_condition_monitoring}.}\label{tab:graz_phases}
\end{table}

To enable direct comparison with the phase by phase degradation trends reported in~\cite{MUTTER2026239640}, temporal evolution within each phase is quantified using the same metric employed in the original study,
\begin{equation}
    \Delta g_t(\log \tau)
    =
    g_t(\log \tau) - g_1(\log \tau),
    \label{eq:drt_diff}
\end{equation}
where $g_1(\log \tau)$ is the first network-estimated \gls{drt} of that phase, serving as the baseline specific to that phase.
While the latent representation introduced in Section~\ref{sec:phase2} offers a better-conditioned basis for tracking system evolution, adopting the original study's own metric here allows the estimated \glspl{drt} to be benchmarked directly against their published results.
This comparison is further strengthened by the constant regularisation of the proposed framework, since all 1248 spectra share the same network weights and $\Delta g_t$ therefore reflects genuine electrochemical change and nothing else.
A method applied per spectrum would impose a different regularisation on each measurement, making it impossible to determine whether differences in the estimated \glspl{drt} reflect genuine electrochemical changes or fitting artefacts.

The evolution of $\Delta g_t(\log\tau)$ tells a coherent degradation story.
Through the first three phases, changes accumulate at the low-frequency end of the distribution: co-electrolysis (P2) adds a new contribution there, and the higher current density of P3 shifts it without amplifying it~\cite[Figure~5a--c]{MUTTER2026239640}, both captured in \figurename~\ref{fig:delta_drt}a--c.
Returning to steam electrolysis (P4) partially undoes this, with the low-frequency contribution retreating and confirming that not all co-electrolysis-induced losses are permanent~\cite[Figure~5d]{MUTTER2026239640} (\figurename~\ref{fig:delta_drt}d).
The most pronounced behaviour occurs during reversible EC/FC operation (P5): the alternating modes leave a strongly periodic imprint on the \gls{drt}, with pronounced high-frequency modulation tied to FC segments~\cite[Figure~5e]{MUTTER2026239640} (\figurename~\ref{fig:delta_drt}e--f).
P6 closes the campaign quietly, with only moderate evolution reflecting the slow pace of ohmic-dominated ageing~\cite[Figure~5f]{MUTTER2026239640}.

Importantly, the same degradation trajectory is independently recovered by the latent space, without any \gls{drt} computation.
Just as in the \gls{sofc} stack monitoring campaign, $D_{10}(t)$~\eqref{eq:latent_distance_general} reduces the entire 2650-hour campaign to a single scalar signal shown in \figurename~\ref{fig:graz_condition_monitoring}, where the slope steepens at the onset of co-electrolysis, relaxes upon recovery, oscillates through the EC/FC phase, and settles into a slow drift in P6, tracing every regime transition identified in the \gls{drt} analysis.

\paragraph{Summary}

\tablename~\ref{tab:findings_comparison} places the key findings of the proposed framework alongside the original results to make the correspondence explicit and verifiable.
Across all three experimental campaigns, spanning different cell types, operating regimes, degradation mechanisms, and campaign durations from 60~h to 3600~h, the proposed framework reproduces the key findings of the original studies without any dataset-specific tuning.
Notably, the framework required neither manual regularisation parameter selection, frequency-band segmentation, equivalent circuit model assumptions, nor dataset-specific architectural modifications.

\begin{table}[h]
    \centering
    \begin{tabular}{@{}p{0.47\textwidth}p{0.47\textwidth}@{}}
        \toprule
        \textbf{Original study} &
        \textbf{This work}                                                                  \\
        \midrule

        \multicolumn{2}{@{}l}{%
        \textit{SOFC stack monitoring campaign~\cite{NUSEV2021,Boskoski2024,Gradisar2026}}} \\

        Fuel-starvation and shutdown events via \gls{uot}~\cite{Gradisar2026}
                                & Peaks in $D_{10}(t)$
        (\figurename~\ref{fig:condition_monitoring})                                        \\

        No prior latent space analysis
                                & Distinct channels per relaxation region
        (\mbox{\figurename~\ref{fig:latent_probe}})                                         \\

        \midrule

        \addlinespace[0.2em]
        \multicolumn{2}{@{}l}{%
        \textit{Operating condition study~\cite{boskoski2024extracting}}}                   \\

        Diffusion onset ${\sim}$20~h, 0.3--2~Hz (Cond.~1)
                                & Latent activation change at ${\sim}$20~h
        (\mbox{\figurename~\ref{fig:conditions123_stacked}a})                               \\

        Air-electrode changes ${\sim}$40~h, 10--40~Hz (Cond.~1)
                                & Latent activation change at ${\sim}$40~h
        (\mbox{\figurename~\ref{fig:conditions123_stacked}a})                               \\

        New equilibrium at ${\sim}$40~h and ${\sim}$70~h (Cond.~2)
                                & Two transitions in latent activation
        (\mbox{\figurename~\ref{fig:conditions123_stacked}b})                               \\

        Pulsating instability throughout (Cond.~3)
                                & Sustained latent activity throughout
        (\mbox{\figurename~\ref{fig:conditions123_stacked}c})                               \\

        \midrule
        \addlinespace[0.8em]
        \multicolumn{2}{@{}l}{%
        \textit{Multi-regime degradation campaign~\cite{MUTTER2026239640}}}                 \\

        Low-freq.\ \gls{drt} grows, case (P2)~\cite[Figure~5b]{MUTTER2026239640}
                                & $\Delta g_t$ increase at low frequencies
        (\mbox{\figurename~\ref{fig:delta_drt}b})                                           \\

        Low-freq.\ \gls{drt} retreats, case (P4)~\cite[Figure~5d]{MUTTER2026239640}
                                & $\Delta g_t$ decrease at low frequencies
        (\mbox{\figurename~\ref{fig:delta_drt}d})                                           \\

        Periodic high-freq.\ modulation, (P5)~\cite[Figure~5e]{MUTTER2026239640}
                                & Periodic \gls{drt} pattern
        (\mbox{\figurename~\ref{fig:delta_drt}e--f})                                        \\

        Moderate evolution, ohmic ageing (P6)~\cite[Figure~5f]{MUTTER2026239640}
                                & Slow latent drift
        (\mbox{\figurename~\ref{fig:graz_condition_monitoring}})                            \\

        \bottomrule
    \end{tabular}
    \caption{Correspondence between findings reported in the original studies and
        signatures identified by the proposed framework. All results in the
        \emph{This work} column were obtained with the same 12{,}000-parameter
        architecture and zero dataset-specific tuning.}\label{tab:findings_comparison}
\end{table}

\FloatBarrier
\section{Conclusion}

This work addressed the problem of estimating the \gls{drt} from \gls{eis} data, an ill-posed inverse task that traditionally relies on careful regularisation or strong modelling assumptions.
We proposed a physics-informed \gls{cnn} that directly predicts the \gls{drt} and reconstructs the corresponding impedance through a discretised formulation of the governing integral relation.
By embedding the physical forward model into the training process, the network is constrained to produce impedance-consistent \glspl{drt} without imposing a fixed number or shape of relaxation peaks, allowing it, in principle, to represent overlapping relaxation processes.

A systematic architectural study showed that physically meaningful reconstructions depend strongly on the imposed inductive bias.
Separating upsampling from convolution in the decoder eliminated checkerboard artifacts, while appropriate choices of bottleneck size and receptive field prevented both over-smoothing and noise amplification.
The resulting 12{,}000-parameter architecture achieves a favourable balance between reconstruction accuracy and generalisation, with range-normalised reconstruction errors below 1.1\% across all three datasets.

What distinguishes this framework from prior approaches is not reconstruction accuracy alone, but what emerges from the shared latent representation.
Because all spectra in a dataset share the same network weights, differences in estimated \glspl{drt} are constrained to reflect genuine electrochemical changes.
This consistency, absent in per-spectrum methods, is what makes the latent space meaningful, as decoder probe analysis revealed that individual channels correspond to distinct relaxation-time regions.
Such a structure emerged solely through optimisation of the physics-informed training objective, directly from the raw \gls{eis} data.

The resulting latent space enables condition monitoring as a direct byproduct.
Euclidean distances in latent space reliably identified every major operational event in a 3600-hour \gls{sofc} campaign, matching conclusions previously obtained through explicit \gls{drt} comparison using \gls{uot}~\cite{Gradisar2026}.

Most significantly, the same architecture was applied without modification to three independent experimental datasets spanning fundamentally different cell types, operating regimes, degradation mechanisms, and campaign durations from 60~h to 2650~h.
In every case, the framework reproduced the key degradation signatures, operational transitions, and \gls{drt} peak evolutions previously established by dedicated per-dataset analyses, including manual regularisation tuning, frequency-band segmentation, expert interpretation, and the frequency-resolved \gls{kl} divergence analysis of the operating condition study.
None of those steps were required here.

Physics-informed learning can therefore replace the manual preprocessing pipeline that has long been the bottleneck in large-scale \gls{eis} diagnostics.
A single, unified, and fully reproducible model is sufficient to extract physically meaningful \glspl{drt}, monitor system health, and generalise across the diverse operating conditions encountered in electrochemical energy conversion research.

\section*{CRediT authorship contribution statement}

\textbf{{\v Z}an Gorenc:}
Conceptualization, Methodology, Software, Validation, Formal analysis,
Investigation, Visualization, Writing -- original draft.

\textbf{{\v Z}iga Gradi{\v s}ar:}
Methodology, Writing -- review and editing.

\textbf{Felix Mütter:}
Investigation, Resources.

\textbf{Vanja Suboti\'{c}:}
Supervision, Investigation, Resources, Funding acquisition.

\textbf{Pavle Bo{\v s}koski:}
Conceptualization, Methodology, Software, Resources, Supervision,
Funding acquisition, Writing -- review and editing.

\section*{Data and code availability}

The source code, pretrained models, reproducibility notebooks, and instructions for downloading the datasets required to reproduce the results presented in this work are publicly available at \url{https://repo.ijs.si/e2pub/cnn_drt.git}.

\section*{Declaration of competing interest}

The authors declare that they have no known competing financial interests or personal relationships that could have appeared to influence the work reported in this paper.

\section*{Acknowledgements}

The authors gratefully acknowledge support for the project ''Probabilistic and explainable data-driven modelling of solid-oxide systems'', jointly financed by the Slovenian Research and Innovation Agency (ARIS), project number J2-4452, and the Austrian Science Fund (FWF), project number I 6251-N. 
The authors alsoacknowledge financial support from the Slovenian Research and Innovation Agency through research programme P2-0001.

\appendix

\setcounter{figure}{0}
\setcounter{table}{0}

\section{Synthetic two-ZARC dataset}\label{app:synthetic_zarc}

The synthetic dataset used for the overlapping-process validation in
Section~\ref{sec:zarc_validation} comprised 1{,}000 two-ZARC impedance spectra.
The ZARC parameters were sampled over the ranges summarised in
\tablename~\ref{tab:synthetic_zarc_ranges}. The resistances and dispersion
parameters were sampled independently from uniform distributions, while the
first characteristic relaxation time was sampled uniformly in logarithmic
space. For the ZARC formulation introduced in
Section~\ref{sec:zarc_validation}, the characteristic relaxation time is
related to the constant-phase-element parameter by

\begin{equation}
    \tau_k
    =
    \left(R_k Q_k\right)^{1/\phi_k}.
\end{equation}

The second relaxation time was determined from a randomly sampled logarithmic
separation according to

\begin{equation}
    \tau_2
    =
    \tau_1 10^{\Delta\log_{10}\tau}.
\end{equation}

\begin{table}[tb]
    \centering
    \begin{tabular}{@{}lll@{}}
        \toprule
        \textbf{Parameter} &
        \textbf{Range} &
        \textbf{Sampling} \\
        \midrule
        $R_{\mathrm{s}}$ [$\Omega$]
            & 5--15
            & Uniform \\
        $R_k$ [$\Omega$], $k=1,2$
            & 30--70
            & Uniform \\
        $\phi_k$, $k=1,2$
            & 0.70--1.00
            & Uniform \\
        $\log_{10}(\tau_1/\mathrm{s})$
            & $-2$--1
            & Uniform \\
        $\Delta\log_{10}\tau$
            & 0.2--2.5
            & Uniform \\
        $\sigma$ [$\Omega$]
            & 0--0.5
            & Uniform \\
        \bottomrule
    \end{tabular}
    \caption{Parameter ranges and sampling distributions used to generate the
    synthetic two-ZARC training dataset.}
    \label{tab:synthetic_zarc_ranges}
\end{table}

Each spectrum was evaluated at 200 logarithmically spaced frequencies between
$10^{-4}$ and $10^{4}$~Hz. Independent zero-mean Gaussian noise was added to
the real and imaginary impedance components,

\begin{equation}
    Z_{\mathrm{noisy}}
    =
    Z
    +
    \epsilon_{\mathrm{re}}
    +
    \mathrm{j}\epsilon_{\mathrm{im}},
    \qquad
    \epsilon_{\mathrm{re}},
    \epsilon_{\mathrm{im}}
    \sim
    \mathcal{N}(0,\sigma^2),
\end{equation}

with $\sigma$ sampled independently for each spectrum from the range reported
in \tablename~\ref{tab:synthetic_zarc_ranges}.

The relaxation-time grid used for physics-based impedance reconstruction
comprised 200 logarithmically spaced points between $10^{-5}$ and $10^{4}$~s.
The dataset was divided into training and validation subsets using an 80:20
split with a fixed random seed. The two test configurations evaluated in
Section~\ref{sec:zarc_validation} were generated separately and excluded from
the training dataset.

\section*{Declaration of generative AI and AI-assisted technologies in the manuscript preparation process}
During the preparation of this work, the authors used Claude to assist with selected coding tasks, code documentation, and streamlining the development workflow. 
Grammarly was subsequently used to review the manuscript for grammar, spelling, and writing style. All outputs generated with the assistance of these tools were critically reviewed, verified, and edited by the authors. The authors take full responsibility for the accuracy, integrity, and final content of the manuscript and the accompanying code.
\FloatBarrier

\bibliographystyle{elsarticle-num-names}
\bibliography{cas-refs}

\end{document}